\documentstyle[twoside,fleqn,espcrc2]{article}

\title{QED: Chiral transition and the issue of triviality }

\author{Aleksandar Koci\' c\\
\vspace{0.25cm}
University of Illinois at Urbana-Champaign, Urbana-IL 61801}

\begin{document}

\begin{abstract}

I give a review and progress report on studies of lattice QED. I
emphasize analytical results and methods that are applied in data analysis.
Also, I derive some bounds for the critical exponents and establish
their connection with scaling violations. Triviality, as realized in
$\phi^4$ theory, is ruled out on theoretical grounds.
I show that the present data,
if analyzed correctly, all lead to the same conclusions. They are
compatible with power law scaling with nongaussian exponents.
\end{abstract}

\maketitle

\section {Introduction}

The most important theoretical
issue related to QED is the question of its
existence and the nature
of the continuum limit. Traditional wisdom is that non-asymptotically
theories do not have an interacting continuum limit in four dimension. This
belief has been based on perturbation theory and, until recently,
hasn't been challenged
outside of the perturbative context. The study of lattice QED is an atempt to
clarify this issue \cite{DKK}\cite{H}\cite{M7a}\cite{Z}.

The central problem in the study of the continuum limit
of a theory is the verification
of hyperscaling \cite{Aiz}.
The starting point is the free energy from which all connected correlation
functions can be generated:
$F_{sing}=t^{2-\alpha}F(h/t^\Delta)$.
Having the dimension of inverse volume, hyperscaling implies that
$F_{sing}\sim \xi^{-d}$.
The renormalized coupling is defined as
$$
g_R=- { {\chi^{(nl)}}\over{\chi^2 \xi^d} }
\eqno(1.1)
$$
where $\chi^{(nl)}=\partial^3 M/\partial h^3$ is the zero-momentum projection
of the connected four-point function. If the action
is gaussian, Wick's theorem applies and the
$N$-point functions factorize. Thus, for a gaussian theory $g_R$ vanishes
and is non-zero otherwise.
Using the hyperscaling hypothesis, eq.(1.1) can be converted into

$$
g_R \sim \xi^{(2\Delta-\gamma-d\nu)/\nu}
\eqno(1.2)
$$
Being dimensionless, $g_R$ should be independent of $\xi$ if $\xi$ is the only
scale. Thus, the validity of hyperscaling
requires that the exponent must vanish. So, hyperscaling
implies the relation, $2\Delta-\gamma-d\nu=0$,
between the critical indices. It is known in
general that the following inequality \cite{J} holds
$2\Delta\leq \gamma+d\nu$.
This means that the exponent  in the expression for $g_R$ is always
non-positive, so that violations of hyperscaling imply that the resulting
theory  is
non-interacting.
Above four dimensions, the exponents are gaussian ($\gamma=1,\Delta=3/2,
\nu=1/2$). In this case, it is easy to verify the above inequality -- it
amounts
to $d\geq 4$. In four dimensions most field theoretical models
critical exponents have
the mean-field values, but with logarithmic corections that drive $g_R$ to
zero. From here, the importance of the knowledge of critical exponents becomes
clear. Establishing their non-gaussian values would be a significant step
towads ruling out the triviality of the theory.

The recent discovery that non-compact $QED$ undergoes a second order
phase transition \cite{DKK}
from a massless to a massive phase with spontaneously
broken chiral symmetry
reopened the old question of the existence of quantum electrodynamics. In what
follows I will survey what has been done.

When studying chiral symmetry breaking in gauge theories one is faced with
the following problem. Because of the singular nature
of the chiral condensate, the chiral
limit is not directly accessible in numerical simulations
and work at finite bare mass is required. In this way
the precise position of the critical coupling is difficult to determine.

\begin{figure}[htb]
\vskip 5truecm
\caption{Mass ratio versus bare mass for $N=2$ theory on a $16^4$
lattice. Solid lines are the fits obtained from the EOS.}
\end{figure}

A usual way to procede is to commit oneself to some scenario
and extrapolate to the chiral limit looking for the
best fit etc.. The index $\delta$ determines the approach to the
chiral limit. It is defined at the critical point by
$m\sim <\bar\psi\psi>^\delta,\,\, (t=0)$.
Here, I will discuss an alternative to the traditional
approach to finding the critical exponents in a given theory
and show how
it is applied to $QED$ where the traditional approach led to ambiguous
and controversial results in the past.

Consider the effect of chiral symmetry breaking on the meson spectrum
from a physical point of view. In the symmetric phase, chiral symmetry
requires the degeneracy between the chiral partners.
Therefore, in the chiral limit
the ratio $R=M^2_\pi/M^2_\sigma =1$.
As the bare mass increases, the ratio
decreases. In the broken phase, however, the ratio vanishes in the chiral
limit because the pion is a Goldstone boson.
Behavior of the mass ratio is displayed in Fig.1. The data are taken from
\cite{N=2}.

The connection between the mass ratio and the order parameter
is obtained from the equation of state (EOS).

$$
h_a=M_a M^{\delta-1}f\bigl({t\over{M^{1/\beta}}}\bigr)
\eqno(1.3)
$$
$$
{  {\chi^{-1}_\sigma }\over{\chi^{-1}_\pi}  } =\delta-
{x\over\beta}{ {f'(x)}\over{f(x)}}
\eqno(1.4)
$$
At the critical point eq.(1.4) results in a simple equation
$$
{ {\chi^{-1}_L}\over{\chi^{-1}_T} }=\delta, \,\,\,\,\,\,\,\, t=0
\eqno(1.5)
$$
The ratio is independent of the symmetry breaking field!
The
knowledge of the mass ratio can be used to determine both the critical point
and
exponent $\delta$ \cite{KKL}.

\section { Two and Four Flavor Theory}

I first discuss the four flavor data.
Simulations in this case were
done on smaller lattices ($12^4$ by DESY \cite{M7} and $10^4$ by Illinois group
\cite{N=4}).
For $N=2$ theory simulations were done on a bigger lattice ($16^4$)
with better statistics by the Illinois group \cite{N=2}.
The quality of the four flavor data
is much more modest and are contaminated by
the finite size effects. This being the case, the conclusions that have been
drawn in the past are to be taken with caution.

\begin{figure}[htb]
\vskip 7truecm
\caption{Mass ratio data versus $m$ taken from the DESY data.
The simbols are explained in the text. }
\end{figure}

The plot of
mass ratios versus the bare mass is
shown in Fig.2. The data are taken from the DESY group \cite{M7}.
Critical parameters from these data are extracted simply.
They are $\beta_c=0.205, \delta=2.3$. The order parameter data,
are shown in Fig.3.
The data on this plot are from the Illinois group \cite{N=4}.
They are consistent
with those of the DESY group. $m=0.03$ was the lowest mass that was
insensitive to the finiteness of the lattice \cite{N=4}. Thus, unlike in the
analysis of the DESY group, $m=0.01$ and 0.02 are
not included here. Both,
$\beta_c$ and $\delta$ obtained this way agree with
the values extracted from
the mass ratios \cite{N=4}.
It should be noted that the critical coupling proposed
by the DESY group \cite{M7}, $\beta_c=0.185$ (dashed line),
lies deeply in the strong coupling
region. Both mass ratio and Fig.3 show this very clearly. No
extrapolations or additional assumptions were made so far.
Note that if $\beta_c$ were
indeed 0.185, as proposed by the DESY group, then the corresponding isotherm
would have to bend upwards. Clearly, such a curvature would correspond to
$<\bar\psi\psi>$ that is too small for a given value of $m$, an effect typical
for symmetry breaking in a small volume.
In the thermodynamic limit, however,
this is not possible on theoretical grounds as I
will discuss later.

\begin{figure}[htb]
\vskip 5truecm
\caption{Order parameter data for the four flavor theory.}
\end{figure}

The EOS (Fig.4)
results in a straight line fit to $f(x)$ and exponent $\beta=0.764$.
This EOS gives predictions for the mass ratio data and the fit is given
by the solid lines  in Fig.2.
The fit is compelling and we
conclude that consistency is satisfactory.

\begin{figure}[htb]
\vskip 5truecm
\caption{EOS fit to power-law scaling for N=4 data from Illinois gruop. }
\end{figure}

Now, I review briefly the atempts of the DESY group to fit the same data to the
logarithmically improved mean field theory.
In ref.\cite{M7} a logarithmically
improved $O(2)$ sigma model is used for fitting purposes and it fits
the data very well.  The EOS reads,
$$
m=
\tau{\sigma\over \ln^p|\sigma^{-1}|}+
\theta{\sigma^3\over \ln|\sigma^{-1}|}
\eqno(2.1)
$$
where $\sigma=\langle\bar\psi\psi\rangle$,
$\tau=\tau_1\theta(1-\beta/\beta_c)$ and $\theta^{-1}
=\theta_o+\theta_1(1-\beta/\beta_c)$.  Choosing specific values for the
five parameters $\beta_c$, $p$, $\tau_1$, $\theta_c$ and $\theta_1$, a
very good fit to the data is found, Fig.5.  The resulting chiral transition
occurs at $\beta_c=0.186(1)$ with mean field critical indices built in.
Since the fit involves five
parameters, its significance is certainly debatable.
Let us subject eq.(2.1) to the same test that the power-law scaling fit
has just passed.  In particular, from Sec.~8 of ref.\cite{M7}
we read off a formula for $R=M^2_\pi/M^2_\sigma$ calculated in the
logarithmically-improved $O(2)$ sigma model. This is done in Fig.6. Obviously,
the fit fails qualitatively.

\begin{figure}[htb]
\vskip 5truecm
\caption{EOS to MF+log scaling, eq.(2.1). DESY data. }
\end{figure}

\begin{figure}[htb]
\vskip 5truecm
\caption{Mass ratio data versus the prediction from the MF+log EOS. Desy data
}
\end{figure}

What is wrong with the MF+log scenario? It
appears to fit the order parameter very well, but it fails with the masses.
The answer is simple. Logarithmic violations of scaling as displayed in
eq.(2.1) can never appear in the case of chiral phase transition.
At the critical point, eq.(2.1) predicts
$$
m\sim {  {<\bar\psi\psi>^3}\over{\log(1/<\bar\psi\psi>)}  }
\eqno(2.2)
$$
Because of the scaling violations,
the RHS vanishes {\it faster} then a pure power. So the "effective" $\delta$
is bigger then its mean-field value.
In the Nambu-Jona-Lasinio model for example
critical EOS gives, on the other hand,
$$
m\sim   <\bar\psi\psi>^3\log(1/<\bar\psi\psi>)
\eqno(2.3)
$$
Unlike for magnets, the log's appear in the numerator --
the RHS vanishes slower then the pure power and the "effective" $\delta$
is smaller then the (pure) mean-field value. As shown in \cite{KK},
this difference in the position of the logarithm i.e. the sign
of the scaling violations is generic for the two models.
It is independent of the approximation and follows
from the basic differences in physics of the two systems.
In \cite{KK} it was shown that exponent $\delta$ satisfies two different
bounds, depending on wheher Goldstone bosons are elementary or
composit. For magnets $\delta>3$, while for fermions $\delta<3$. These
bounds are respected by the logarithmic violation of scaling in four
dimensions \cite{KK}. The origin of these bounds on $\delta$ lies in the
fact that for magnets weak coupling phase is broken wheras for fermions
the situation is reversed.
Thus, the failure of the fits of ref.\cite{M7} is understandable and expected
since it is borrowed from the magnetic context and applied to the fermions.
The source of errors in the fits
of ref.\cite{M7} is missidentifiacion
of the physics and the lack of control of finite size effects.
Because of the small ($12^4$)
lattice, $\beta=0.185$ that was mistakenly identified
with the critical coupling
by the DESY group was deep at the strong coupling phase.
Such a misidentification coresponds to a smaller size of
the strong coupling phase and the larger value of $\delta$.

Concerning the other results reported by the DESY group in \cite{M7},
several comments are in order. Some preliminary
attempts to extract renormalized charge and
fermion mass have been made in ref.\cite{M7}.
I should mention some obvious technical problems
that concern
the state of affairs related to this attempt. Like order parameter studies,
simulations of the renormalized charge and fermion mass were done on a
$12^4$ lattice. The procedure that was employed assumed the masslessness
of the photon. The renormalized charge was then extracted as the residue of
the photon propagator evaluated at zero momentum. However, due to the finite
volume constraint, the propagator was evaluated at finite momentum and the
result was extrapolated to $k=0$. This is a mysterious step since ref.\cite{M7}
does not make any attempt to clarify the origin of the extrapolation procedure
and its stability to a change in volume.

In fact, this
size of the lattice was too small to yield qualitatively correct results
for the quantities like $<\bar\psi\psi>$ and meson masses. These data were
seriously distorted by the finiteness of the volume.
The photons, being presumably massless,
can only be more sensitive to such a small volume. The same holds for the
fermion mass since fermions are charged and are, in principle, much more
sensitive to the presence of the massless photons then are mesons.
Therefore, none of the results and the conclusions regarding the photon
and fermion propagator, can be taken seriously.
The quality of the fermion mass and renormalized charge data is not good
to yield any quantitative statements and, if taken seriously, they lead
to incorrect and inconsistent result.

For example, consider the fermion mass data taken from the DESY group.
In general, for any mass, $M_A$, in the scaling region the following
equation holds: $M_A=t^\nu G_A(m/t^\Delta)$.
Thus, the ratio $M_F/M_\sigma =G(m/t^\Delta)$ is a function of only one
variable. This is true if both masses scale.
The data for $M_F/M_\sigma$, taken from \cite{M7},
are shown in Fig.7.
Regardless of the value of $m$, the ratio at the critical point,
$t=0$, always has the same value. Thus, the curves $R$ vs $\beta$ have to
cross at $\beta_c$ (insert of Fig.7).
However, data show that {\it the curves never cross!}
The fermion mass, as taken
from the data,  does not scale. It doesn't show proper restoration of
chiral symmetry probably because finite volume effects keeps it too heavy.
Clearly, if scaling properties were to be recovered, the extrapolation would
drive us to the weaker coupling in order to reduce the fermion mass to its
value compatible with the phase of restored symmetry. This is why the
conclusions obtained through extrapolation of the data in ref.\cite{M7}
agreeed so
well with renormalized perturbation theory -- high fermion mass was confused
for low renormalized charge. This is how the authors of ref.\cite{M7}
misinterpreted their data as supporting the zero charge continuum limit.

\begin{figure}[htb]
\vskip 5truecm
\caption{Fermion-sigma mass ratio versus bare coupling. DESY data.
The insert is an illustration of the behavior of mass ratio
in the scaling region.}
\end{figure}

As far as the two flavor data is concerned, the extensive studies have been
done on lattices ranging from $10^4$ to $16^4$. The results have been
reported in ref.\cite{N=2}. The same strategy has been applied
as in the case of four flavor theory. In Fig.1. the
mass ratio data are confronted
with the predictions of the power-law EOS (solid lines).

A different approach, using microcanonical method,
is adopted by the Zaragoza group \cite{Z}.
The results for the
critical coupling and the critical indices obtained by the Illinois and
Zaragoza group agree.

A brief comment about perturbation theory should be made.
The applicability of perturbation theory in the
case of scalar theories has been diagnosed early on.
For example, in the $\sigma$-model the weak
coupling phase is at low-temperatures. Due to the presence
of Goldstone bosons, all the correlation functions are saturated with the
massles states and the entire low-temperature phase is massless. Every point
is a critical point in the limit of vanishing magnetic field.
Thus, the
low temperature expansion is an expansion in powers of $T$.
The terms of the form $\exp(-M/T)$
are absent and there is no danger that they will be omitted by using
perturbation theory.
In this way, in principle, the critical region can be accessed through
perturbation theory which is quite accurate close to two dimensions because
the critical temperature moves to the origin.
Clearly such reasoning can not be applied to fermions simply because
the weak coupling phase is symmetric. Thus, no matter how small the
coupling is, perturbation theory omits the Goldstone physics as a matter of
principle. It can not produce bound states that accompany the chiral transition
and  its use is questionable in this case.

This holds for $QED$ as well although renormalized charge is bounded
from above \cite{L}. In fact, it is difficult to see how
perturbation theory  could communicate with the
chiral transition. For example, the
fact that the value of the mass ratio at the critical point
is independent of the bare mass is a truely nonperturbative result and
could not be obtained from any expansion in powers of
$\alpha_R$. Since
$\alpha_R$ is dimensionless, at the critical point it must be independent
of $m$ as well, which is clearly impossible in perturbation theory where
screening is sensitive to the fermion mass. In other words, conflicting RG
trajectories {\it must}
appear if perturbative analysis of the renormalized charge is used.

\section {Conclusions}

We have seen so far that the problem of analyzing the scaling region of
an unknown theory is not straightforward. Depending on the accuracy of the
data, ambiguous results can follow. If $\beta_c$ is underestimated,
and appropriate extrapolations done, the "best fit" is given by mean
field exponents supplemented with the log's in the wrong place! We have been
able to eliminate this possibility on theoretical grounds \cite{KK}.
At this stage it is fair to say only that the data {\it support} the power-law
scaling with non-gaussian exponents.
The data of all three groups\cite{N=4}\cite{M7}\cite{Z} agree on this
point.
Triviality as realized
in $\phi^4$ theory is definitely ruled out.
The scenario proposed by the DESY group \cite{M7} is dismissed on
theoretical grounds, without a need to consult the data.
It remains to be seen whether the power law scaling with $\delta\approx2.3$
can be distinguished numerically from
mean-field plus log's with the log's as in Nambu-Jona-Lasinio model.
This is the question that needs to be addressed
eventually since the value of $\delta$ obtained from the simulations
and is not too far from its
mean-field value. Work on this problem is in progress.

\centerline{\bf Acknowledgement}
Most of the work reported here was done in collaboration with John Kogut
and Maria-Paola Lombardo.
I wish to acknowledge the discussions with
V. Azcoiti, E. Fradkin, P. Hasenfratz, E. Seiler
and P. Weisz. This work is supported by NSF-PHY 92-00148.


\begin{thebibliography}{9}

\bibitem{DKK} E. Dagotto, A. Koci\' c and J. B. Kogut, {Phys. Rev. Lett.}.
{\bf 60}, 772 (1988); {\bf 61}, 2416 (1988).

\bibitem{H}  A.~M.~Horowitz, Phys.Lett. {\bf B219} (1989) 329.


\bibitem{M7a}  M. Gockeler, et al.,
{Nuc. Phys.} {\bf B334}, 527 (1990).


\bibitem{Z}  A. Azcoiti, G. Di~Carlo and A. F. Grillo,
Int. Jour. Mod. Phys. {\bf A8}, 2239 (1993).


\bibitem{Aiz}  M. Aizenman, Comm. Math. Phys.
{\bf 86}, 1 (1982); C. Aragao de Carvalho, S. Caracciolo and J. Frohlich,
Nucl. Phys. {\bf 215}[FS7], 209 (1983).

\bibitem{J}  B.~D.~Josephson, Proc. Phys. Soc. {\bf 92} (1967) 269, 276.


\bibitem{M7}  M. Gockeler, et al., {Nuc. Phys.} {\bf B371}, 713 (1992).

\bibitem{KKL}  A. Koci\' c, J. B. Kogut and M.-P. Lombardo,
{Nuc. Phys.} {\bf B398}, 376 (1993).

\bibitem{N=2}  S.~J.~Hands, et al., CERN-TH 6609/92.

\bibitem{N=4}  A.~Koci\'{c}, J.~B.~Kogut and K.~C.~Wang, Nucl. Phys. {\bf B398}
(1993) 405.

\bibitem{L}  M. L\" uscher, Nucl. Phys. {\bf B341}, 341 (1990).

\bibitem{KK}  A.~Koci\'{c} and  J.~B.~Kogut, {\it Compositness, Triviality
and Bounds on Critical Exponents for Fermions and magnets}, Illinois preprint.


\end{thebibliography}
\end{document}